\documentclass[5p,sort&compress]{elsarticle}

\usepackage{lineno,hyperref}
\modulolinenumbers[5]

\usepackage[utf8]{inputenc}
\usepackage{amsmath}
\usepackage{amsfonts}
\usepackage{amssymb}
\usepackage{graphicx}
\usepackage{caption}

\usepackage[separate-uncertainty=true]{siunitx}

\title{Monitoring of laser metal deposition height by means of \mbox{coaxial laser triangulation}}
\begin{document}

\begin{frontmatter}
	
\author[mymainaddress]{Simone Donadello\corref{mycorrespondingauthor}}
\cortext[mycorrespondingauthor]{Corresponding author}
\ead{simone.donadello@polimi.it}

\author[mymainaddress]{Maurizio Motta}
\author[mymainaddress]{Ali G\"{o}khan Demir}
\author[mymainaddress]{Barbara Previtali}

\address[mymainaddress]{Department of Mechanical Engineering, Politecnico di Milano, Via La Masa 1, 20156 Milan, Italy}

\begin{abstract}
	Laser metal deposition (LMD) is an additive manufacturing technique, whose performances can be influenced by several factors and parameters. Monitoring their evolution allows for a better comprehension and control of the process, hence enhancing the deposition quality. In particular, the deposition height is an important variable that, if it does not match the process growth, can bring to defects and geometrical inaccuracies in the deposited structures. The current work presents a system based on optical triangulation for the height monitoring, implemented on a LMD setup composed of a fiber laser, a deposition head, and an anthropomorphic robot. Its coaxial and non-intrusive configuration allows for flexibility in the deposition strategy and direction. A measurement laser beam is launched through the powder nozzle and hits the melt pool. A coaxial camera acquires the probe spot, whose position is converted to relative height. The device has been demonstrated for monitoring the deposition of a stainless steel cylinder. The measurements allowed to reconstruct a spatial map of the height variation, highlighting a transient in the deposition growth which can be explained in terms of a self-regulating mechanism for the layer thickness.
\end{abstract}


\end{frontmatter}
\section{Introduction}
Additive manufacturing has gained interest in many research and industrial fields, from aerospace to biomedical applications, introducing big advantages in terms of flexibility for the design and direct realization of solid objects with complex and custom geometries \cite{frazier_metal_2014}. Within such context, the laser metal deposition (LMD) process consists in melting a metallic powder by means of the thermal energy provided by a high-power laser beam. Typically the powder is carried by an inert gas and sprayed by a nozzle, with a coaxial laser beam passing through the nozzle and overlapping with the powder flow, hence generating a melt material pool on a substrate. A solid layer is obtained along the deposition track after the material solidification, and three-dimensional (3D) structures can be build by repeating the procedure over the previous layers. 

The LMD process depends on several parameters, including the laser power, the deposition speed, and the powder flow rate. Moreover, the deposition can be influenced by physical quantities which can vary during the process, such as the substrate temperature. In fact, if the temperature changes due to unbalance between heat accumulation and conduction, the powder melting can be eventually favored, introducing variability in the deposition growth which can lead to the formation of defects or irregularities in the deposited structure. The quality requirements in production environments and the high costs of additive manufacturing encourage the development of specific feedback systems for the adaptive control of the process parameters \cite{mazumder_closed_2000,purtonen_monitoring_2014,shamsaei_overview_2015,sammons_repetitive_2018}. For this reason several aspects of the deposition process have been monitored and studied with different techniques, such as pyrometers or camera vision systems for measuring the substrate temperature or the deposition growth, as reported in many research works \cite{meriaudeau_control_1996,hu_sensing_2003,hofman_camera_2012,tapia_review_2014,everton_review_2016,kim_review_2018}.

The distance between the nozzle and the substrate, called standoff distance (SOD), is another important parameter of the LMD process. As a matter of fact, the deposition rate is strongly influenced by the deposition height, since the latter determines the overlapping factor between the focused laser beam and the convergent powder flow \cite{peyre_analytical_2008,kovalev_theoretical_2011}. If the SOD departs from its optimal value, the powder-laser interaction can be altered, resulting in process growth variations and, consequently, reduced deposition quality and geometrical inaccuracies \cite{pinkerton_significance_2004,zhu_influence_2012}.

Several approaches for studying the deposition growth in LMD and laser cladding can be found in literature. Firstly, the deposition height can be included in models developed from numerical simulations \cite{labudovic_three_2003,amine_investigation_2014,manvatkar_spatial_2015}. These allow to explore generic geometrical configurations and process parameters, with the main drawback of a high computational cost which must be carried out offline. Otherwise, an indirect control of the process parameters can be obtained by correlating the height information to other physical quantities, which can be deduced by analyzing the melt pool images acquired with cameras during the process \cite{fathi_clad_2007,bi_development_2007}. However, such kind of method might not be robust against variations of the process parameters, requiring the development of specific, and possibly complex, semi-empirical models.

The deposition height can be also extracted from the 3D reconstruction of the deposited object, e.g., obtained by means correlation analysis algorithms for images taken with off-axis cameras from one or more points of view as reported by several studies \cite{hua_feedback_2005,asselin_development_2005,iravani-tabrizipour_image-based_2007,song_control_2012,biegler_-situ_2018}. These approaches give rich information and seem to be suitable for research purposes, while their usage for production applications might be limited by the complexity of the monitoring setups, whose size might also obstruct the movements of the deposition equipment.

A wide class of optical methods for the deposition growth monitoring is based on the triangulation principle. Its classical implementation exploits a tilted laser beam probing the target surface, with an image sensor used for detecting the probe spot position. A custom configuration of such working principle was previously demonstrated on a wire-LMD setup for the process control \cite{heralic_increased_2010}. Subsequent works reported the usage of commercial laser displacement sensors \cite{tang_layer--layer_2011,denlinger_effect_2015,heigel_situ_2015,segerstark_investigation_2017} or 3D scanners \cite{heralic_height_2012}. In general, these kinds of high-precision instruments are characterized by a nominal resolution of few micrometers, allowing very accurate measurements during the process. Their intrinsic limits are mainly related to the off-axis arrangement of the probe beam. In fact, this may introduce anisotropy in the measurement direction, possibly suffering of blind zones or shadowing effects, hence limiting the flexibility of their application, especially in the case of deposition of complex and big geometries.

The method presented in the current work reinterprets and simplifies the common triangulation implementation for monitoring the deposition height during the LMD process, introducing several advantages. In fact, the proposed system represents a simple, non-intrusive, and cheap solution for monitoring inline the deposition growth, as well as a non-destructive diagnosis tool of the deposited structure \cite{donadello_coaxial_2018}. The device for the in situ height measurement has been integrated on a setup composed of a fiber laser and a robotized deposition head. The coaxial configuration of the probe laser beam shares the optical path of the high-power laser within the deposition head, allowing for flexibility in the deposition strategy and being independent on the direction of the transverse movements. The direct measurement of the melt pool distance does not require the development of process models, which may depend on the deposition parameters and materials.

The triangulation system has been operated while building a multi-layer hollow cylinder from stainless steel powder, demonstrating its robustness against the direction of the robot movements. The results highlighted a self-regulating mechanism in the layer thickness. The latter, after an initial transient, tends to an equilibrium condition, interpreted as a result of compensation between concurrent thermal and powder defocusing effects. A 3D spatial reconstruction obtained from the measurements allowed to visualize structural defects of the deposited cylinder along the growth direction. Although the sensitivity of the proposed method might be lower if compared to some of the optical instruments utilized in the studies cited before, this is sufficient for many applications, such as the detection of sub-millimeter height mismatches or the implementation of closed-loop feedback controllers on the actual layer thickness, while flexibility in the measurement is gained due to the coaxial configuration. Finally, the examined monitoring device is composed by simple and low-cost components, which are promising factors for its usage even in industrial environments, with minimal changes to existing setups.

\begin{table}[ht]
	\label{tab:nomenclature}
	\begin{center}       
		\begin{tabular}{ll}
			\hline
			\multicolumn{2}{l}{Nomenclature} \\
			\hline
			SOD & standoff distance \\
			$H$ & height of the deposited structure \\
			$h$ & thickness of a single deposited layer\\
			$D$ & height programmed to the robot \\
			$d$ & incremental height programmed to the robot \\
			$z_1$ & distance between focal plane and target \\
			$z_0$ & initial value of $z_1$ at the reference SOD \\
			$\Delta z$ & relative height with respect to the reference $z_0$ \\
			$y_1$ & probe spot position in the target plane \\
			$y_2$ & probe spot coordinate on the camera sensor \\
			\hline
		\end{tabular}
	\end{center}
\end{table} 
%\clearpage

\section{System design and implementation}
\subsection{LMD setup}
\begin{figure}[ht]
	\centering
	\includegraphics[width=0.85\linewidth]{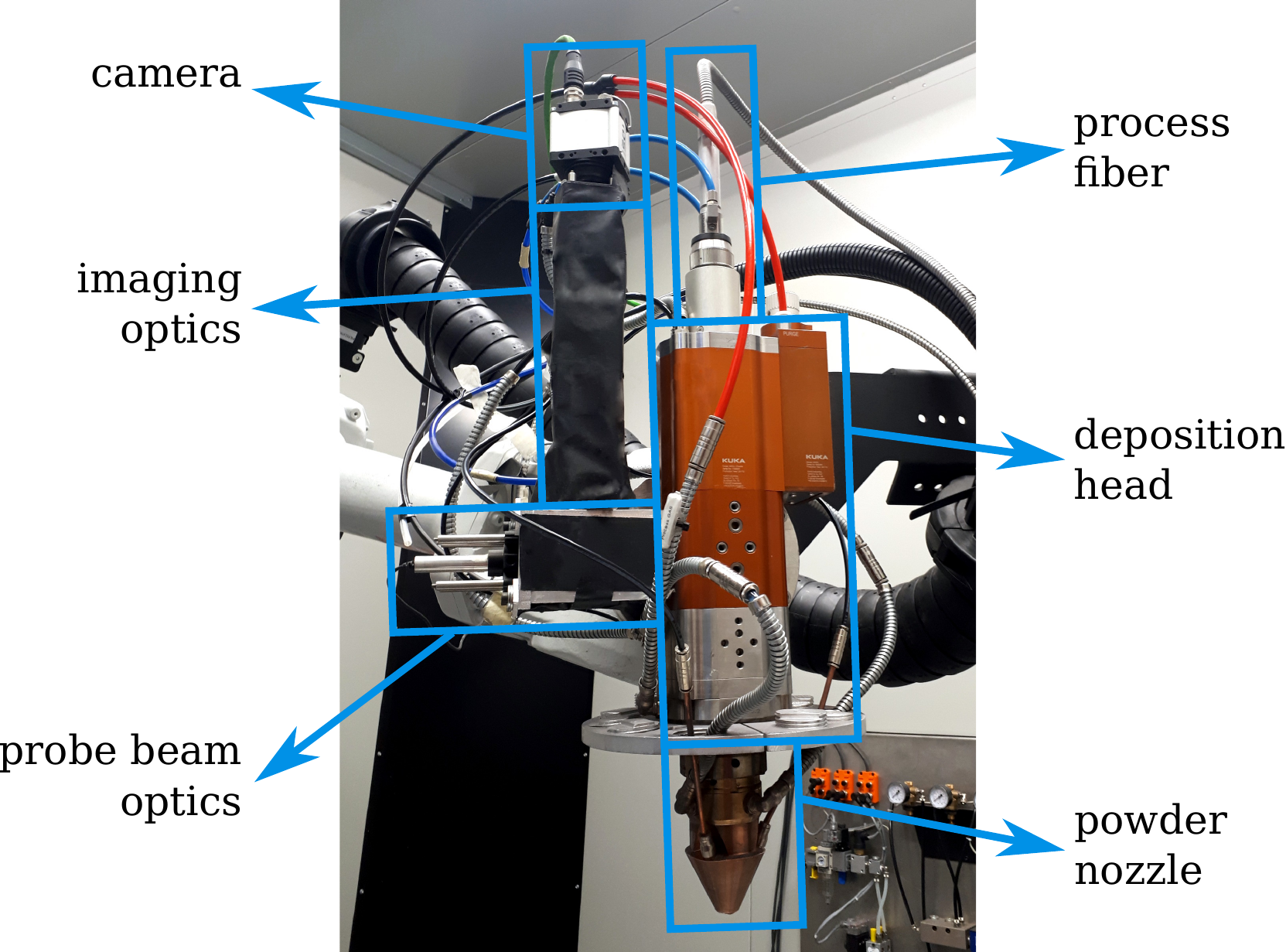}
	\caption{Side view of the experimental setup for the height monitoring during the LMD process.}
	\label{fig:photo-setup}
\end{figure}

The monitoring system has been implemented on the LMD setup illustrated in Fig.~\ref{fig:photo-setup}, whose characteristics are summarized in Table~\ref{tab:lmd-parameters}. The equipment is based on a deposition head (\textsc{Kuka Reis MWO-I}) mounted on a 6-axis anthropomorphic robot (\textsc{ABB IRB 4600-45}). The optical energy source is a \SI{1070}{nm} active fiber laser (\textsc{IPG YLS-3000}) having \SI{3}{kW} maximum power. The \SI{50}{\micro\meter} feeding fiber of the laser is connected to the \SI{400}{\micro\meter} process fiber through a fiber-to-fiber coupler, delivering the optical radiation to the deposition head. The process laser beam is collimated with a \SI{129}{mm} lens, then it gets focalized toward the deposition region by the lens $L_1$ with focal length $f_1=\SI{200}{mm}$.

\begin{table}[ht]
	\centering
	\caption{Characteristics of the LMD setup.}
	\label{tab:lmd-parameters}
		\begin{tabular}{ll}
			\hline 
			Process laser source & \textsc{IPG YLS-3000} \\ 
			Maximum laser power & \SI{3}{\kilo\watt} \\ 
			Laser emission wavelength & \SI{1070}{nm} \\ 
			Deposition head & \textsc{Kuka Reis MWO-I} \\ 
			Anthropomorphic robot & \textsc{ABB IRB 4600-45} \\ 
			Powder nozzle & \textsc{Fraunhofer ILT} \\ 
			& \textsc{3-JET-SO16-S} \\
			Reference standoff distance & \SI{12}{mm} \\ 
			Process lens focal length & $f_1=\SI{200}{mm}$\\
			\hline 
		\end{tabular} 
\end{table}

The metallic powder to be deposited is fed to the three-jet powder nozzle (\textsc{Fraunhofer ILT 3-JET-SO16-S}) of the deposition head by a powder feeder (\textsc{GTV Twin PF 2/2-MF}), using nitrogen both as vector and nozzle shielding gas. The powder is ejected by three orifices configured at \ang{120} from each other, converging to the deposition zone and generating a powder cone. The standoff distance between the nozzle tip and the substrate is set to the reference value of \SI{12}{mm} at the beginning of the process.

\subsection{Coaxial triangulation setup}
\begin{figure}[ht]
	\centering
	\includegraphics[width=0.95\linewidth]{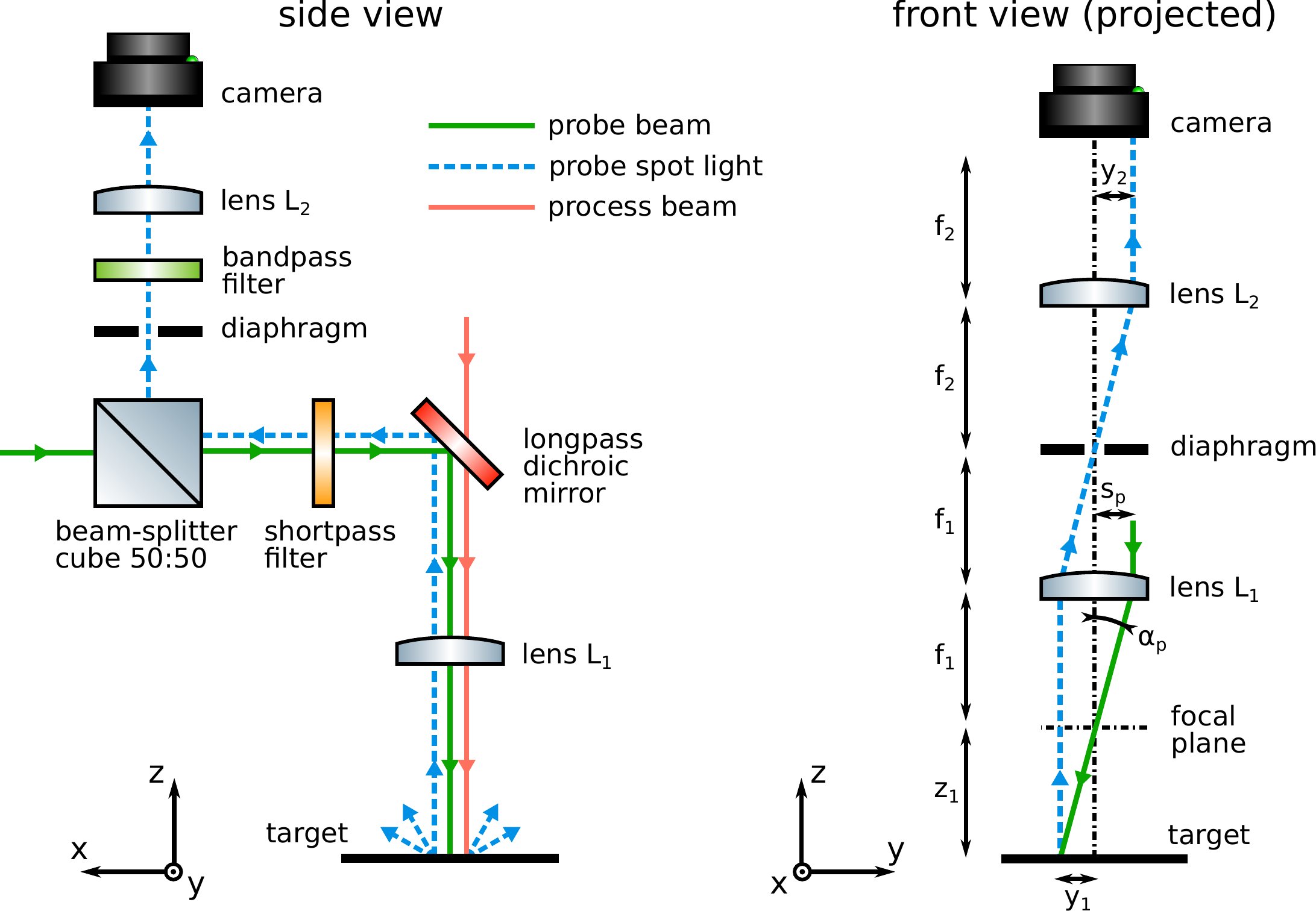}
	\caption{Sketch of the optical setup used for the height monitoring. On the left, side view of the probe beam path, with the imaging system for the detection of the probe spot on the target. On the right, projected and simplified front view, with the main dimensions involved in the triangulation measurement.}
	\label{fig:setup-front-side}
\end{figure}

\begin{figure}[ht]
	\centering
	\includegraphics[width=0.6\linewidth]{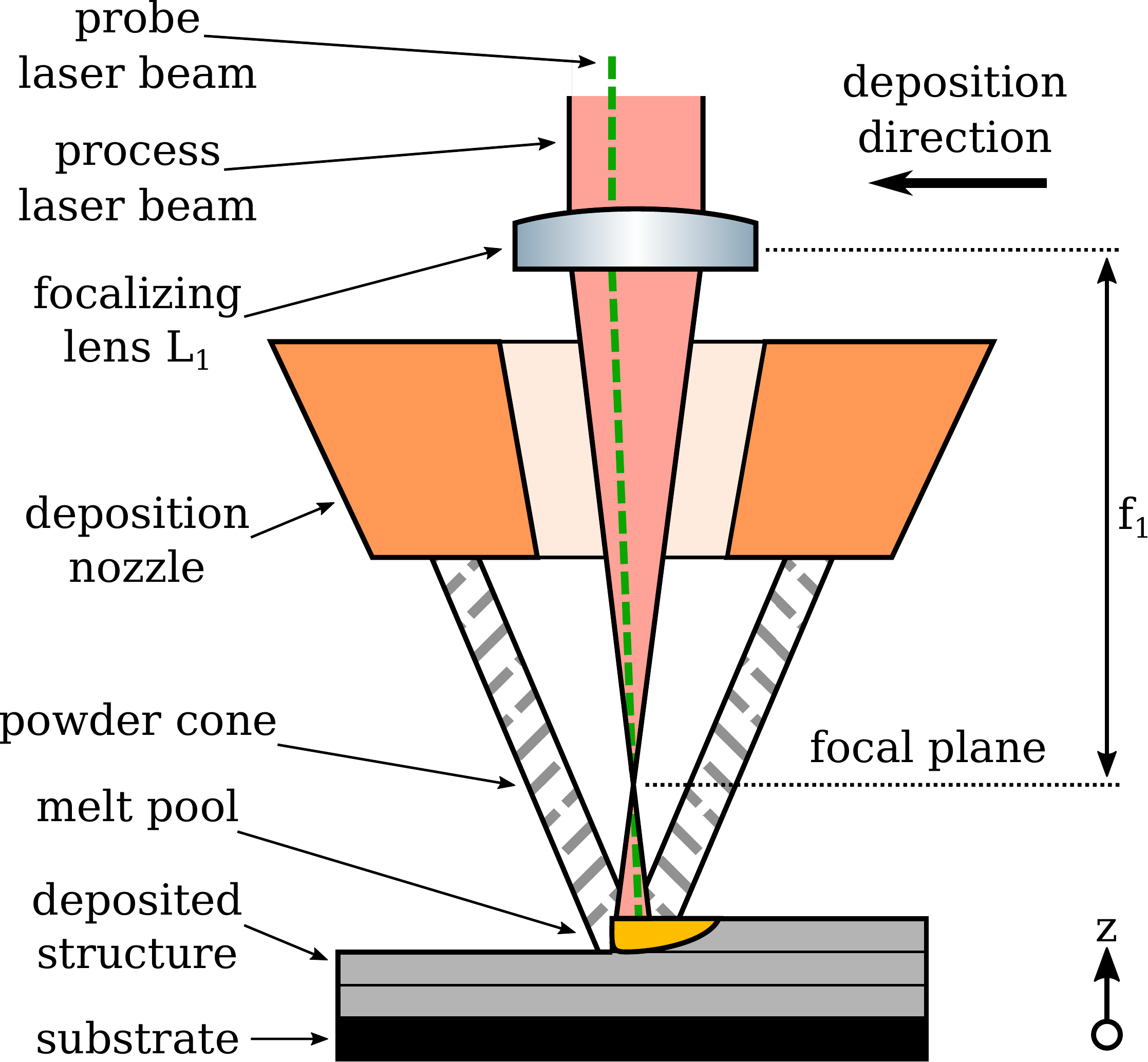}
	\caption{Schematic representation of the powder nozzle during the deposition, with the measurement beam probing the melt pool.}
	\label{fig:setup-ded}
\end{figure}

The setup for the deposition height monitoring includes a probe laser beam and a coaxial imaging system, both housed in a custom unit attached sideways to the deposition head as illustrated in Fig.~\ref{fig:photo-setup}. The optical chain of the triangulation device partially shares the process beam path as sketched in Fig.~\ref{fig:setup-front-side}, probing the deposition area as represented in Fig.~\ref{fig:setup-ded}. The component characteristics of the measurement apparatus are summarized in Table~\ref{tab:triangulation-parameters}.

\begin{table}[ht]
	\centering
		\caption{Characteristics of the triangulation setup.}
		\label{tab:triangulation-parameters}
		\begin{tabular}{ll}
			\hline 
			Probe laser source & \textsc{Thorlabs CPS532} \\
			Laser power & \SI{4.5}{\milli\watt} \\
			Emission wavelength & \SI{532}{nm} \\
			Collimated beam diameter & \SI{3.5}{mm} \\
			Imaging lens focal length & $f_2 = \SI{125}{mm}$ \\
			Focused spot diameter & $\SI{200}{\micro \meter}$ \\
			CCD camera & \textsc{IDS UI-6230RE-M-GL} \\
			Image magnification & $M\simeq 0.62$ \\
			Camera pixel size & $p_s=\SI{4.65}{\micro\meter}$ \\
			Acquisition frame rate & \SI{98.4}{\hertz} \\ 
			\hline 
		\end{tabular} 
\end{table}

The probe source is a laser diode module (\textsc{Thorlabs CPS532}) emitting \SI{4.5}{\milli\watt} at \SI{532}{\nano\meter}. The collimated beam shape is circular, with a \SI{3.5}{\milli\meter} diameter. The probe beam passes through a $50$:$50$ non-polarizing beam-splitter cube (\textsc{Thorlabs BS004}), exploited for superimposing the mutually orthogonal probe and imaging optical axes. Half of the beam intensity is transmitted by the cube and continues toward the deposition head, half is reflected and lost. The probe and process beams are superimposed by means of a dichroic mirror, tilted at \ang{45} relatively to the optical axis of the deposition head. The dichroic mirror transmits the infrared process radiation and reflects about the $72\%$ of the green probe light. Then the probe beam passes through the convergent lens $L_1$ of the deposition head having nominal focal length $f_1=\SI{200}{mm}$, whose focal plane is about \SI{3}{mm} out from the nozzle tip. The beam exits from the deposition head by crossing the \SI{6}{mm} diameter aperture of the powder nozzle.

In the case that a diffusive target is present out from the deposition head, the probe beam gets scattered from the incidence point on the target surface. From a merely geometrical consideration and in the assumption of isotropic scattering, only a small fraction of the diffused light can go backward through the nozzle aperture. Such scattered light is collected by the lens $L_1$ of the deposition head, then it gets deflected by the dichroic mirror toward the triangulation unit. Half of the scattered light is transmitted and lost by the beam-splitter cube, while half is reflected to the imaging arm of the setup. A second convergent lens $L_2$ with focal length equal to $f_2=\SI{125}{mm}$ is placed at the optical distance $f_1+f_2$ from $L_1$, in a telescope configuration whose magnification $M$ is limited by the process lens of the existing LMD setup:
\begin{equation}
\label{eq:magnification}
M = \frac{f_2}{f_1} \simeq 0.62\,.
\end{equation}
The probe spot image is acquired with a CCD mono\-chrome camera (\textsc{IDS UI-6230RE-M-GL}) placed at distance $f_2$ from the imaging lens $L_2$. The camera sensor has a maximum resolution of \num{0.8e6} pixels, with pixel size equal to $p_s=\SI{4.65}{\micro\meter}$. The acquisition is performed on a cropped area of $340\times120$ pixels. The integration time is \SI{10}{\milli\second}, with the \SI{98.4}{\hertz} frame rate setting the measurement temporal resolution. The limit for the lateral spatial accuracy is determined by the probe spot diameter, which is measured as about \SI{200}{\micro\meter} around the focal point.

Several factors may disturb the measurement while depositing. First of all, the infrared light of the high-power process laser might be partially scattered back from the deposition area or from parasitic reflections, reaching the monitoring setup. Such process radiation is extinguished by a shortpass wavelength filter (\textsc{Thorlabs FESH1000}) with \SI{1000}{nm} cutoff, placed at the interface between the triangulation unit and the deposition head. The presence of other off-focus light beams, e.g., given by back-reflections of the probe light from the head optical elements, might also hide the probe signal on the target. These are suppressed by a spatial filter, obtained with a \SI{1}{\milli\meter} diameter diaphragm placed in the common focus of the $L_1$ and $L_2$ lenses. Another important noise source during the process is represented by the broadband thermal emission from the melt pool, as well as the ambient light. Even if the typical thermal emission is not strong at the probe wavelength, the usage of a notch spectral filter (\textsc{Thorlabs FL05532-1}) with a \SI{1}{nm} FWHM bandwidth around \SI{532}{nm} allows to enhance the signal-to-noise ratio for the weak signals of the scattered probe light.

Within the other possible disturbances, it must be considered that the probe beam overlaps and crosses the powder which flows from the nozzle. However, as showed by some preliminary tests, the scattering losses introduced by the powder are negligible. Perturbations from the process-induced plume can be also neglected since, due to the relatively low beam irradiance, the generation of plasma or plume in LMD is limited if compared to other processes such as laser cutting, welding, or ablation. Finally, the probe spot detection might partially suffer from mutable speckle patterns and variations in the target reflectivity, although the height measurement is actually related to the spot position instead of the absolute spot intensity.

\subsection{Measurement principle}
If the collimated probe beam is perfectly aligned with the optical axis of the deposition head, hence hitting the lens $L_1$ in its center, the focused beam will be coaxial to the process beam. Conversely, if the probe beam is translated by $s_p$ from the lens center, it will be deflected by an angle $\alpha_p$ relatively to the optical axis. By neglecting the beam size and with an orthogonal incidence on the lens, the probe beam can be approximated as a single ray passing through the lens focal point. As it can be deduced from the sketch of Fig.~\ref{fig:setup-front-side}, the beam deflection angle $\alpha_p$ is equal to
\begin{equation}
\alpha_p = \tan^{-1}\left(\frac{s_p}{f_1}\right)\,.
\end{equation}

The probe beam deflection introduced by the deposition head lens is the basis of the triangulation measurement. In fact, with $\alpha_p>0$, the position of the probe spot on the horizontal target surface varies depending on the vertical height. From simple geometrical optics considerations, the coordinate $y_1$ of the probe spot in the target plane, defined along the deflection plane and referred to the optical axis, is proportional to the distance $z_1$ between the focal plane and the target itself:
\begin{equation}
y_1 = \frac{s_p}{f_1}z_1\,.
\end{equation}
Taking into account the magnification $M$ of Eq.~\eqref{eq:magnification}, the position $y_2$ of probe spot image on the camera sensor plane is
\begin{equation}
\label{eq:y_2}
y_2 = My_1 = \frac{f_2 s_p}{f_1^2}z_1\,.
\end{equation}

During the process the target is the melt pool on the deposited structure, whose relative distance varies with time depending on the deposition growth and on the focal plane position, the latter being joined with the deposition head. It is convenient to introduce the relative height $\Delta z$, defined as
\begin{equation}
\label{eq:delta-zeta}
\Delta z = z_1 - z_0\,,
\end{equation}
where $z_0$ is the reference distance measured at the beginning of the deposition, i.e., when the nozzle is at the nominal SOD.

A higher probe beam offset $s_p$, hence a higher deflection angle $\alpha_p$, would be desirable for a better measurement sensitivity. However, the requirement of independence on the deposition direction introduces the constrain for a quasi-coaxial configuration, fulfilled when $\alpha_p$ is sufficiently small. This means that, considering a reasonable measurement range, the probe spot must remain within the deposition region, specifically the melt pool, whose width is typically of the order of \SI{1}{mm}. Therefore, with $y_1$ ranging as $\pm \SI{0.5}{mm}$ and for a $\Delta z$ range of $\pm \SI{10}{mm}$, the beam offset must be $s_p < \SI{10}{mm}$. The clear aperture of deposition head interface limits $s_p$ to a maximum of about \SI{8}{mm}. The \SI{6}{mm} nozzle aperture is sufficiently wide and does not introduce further limitations to the probe beam passage.

The acquired images of the probe spot are analyzed with a \textsc{Python} code, which integrates each image along the axis that is insensitive to height variations, removes the background baseline, and extracts the probe spot position by fitting the intensity profile to a one-dimensional Gaussian function. The spot center coordinate, labeled as $y_2'$, is measured in pixel number and interpolated on a sub-pixel scale. It must be noted that the value of $y_2'$ is referred to the CCD sensor origin, i.e., it is equal to $p_sy_2$ plus an offset, with $p_s$ being the pixel size. Therefore, the relation between $z_1$ and $y_2$ of Eq.~\eqref{eq:y_2} can be rewritten in terms of relative height $\Delta z$ and camera coordinate $y_2'$ as
\begin{equation}
\label{eq:delta-zeta-coeff}
y_2' = \beta_0 + \beta_1 \Delta z\,,
\end{equation}
with $\beta_0$ a constant term which depends on $z_0$ and on the optical alignment. The proportionality factor
\begin{equation}
\beta_1 = \frac{f_2 s_p}{f_1^2p_s}
\end{equation}
defines the estimated vertical sensitivity of the triangulation measurement, although the continuous fitting procedure overcomes the pixel discretization limit.

The system has been calibrated by measuring $y_2'$ as a function of known values of $\Delta z$, i.e., controlling the position of a dummy target with a precision vertical translation stage \cite{donadello_coaxial_2018}. The coefficients of Eq.~\eqref{eq:delta-zeta-coeff} have been extracted with linear regression from the calibration data, and they are reported in Table~\ref{tab:triangulation-calibration}. In the actual experimental conditions, the $\Delta z$ variation corresponding to a single camera pixel is equal to
\begin{equation}
\frac{1}{\beta_1} = \SI{0.271 \pm 0.003}{mm \per pixel}\,.
\end{equation}
The expected off-axis displacement $s_p$ and the deflection angle $\alpha_p$ can be calculated from $\beta_1$ as \SI{5.6}{mm} and \ang{1.6} respectively, considering the nominal focal lengths $f_1$ and~$f_2$.

\begin{table}[ht]
	\centering
		\caption{Characteristics of the probe beam geometrical configuration, calculated from the calibration coefficients referring to the linear relation between $\Delta z$ and $y_2'$ defined in Eq.~\eqref{eq:delta-zeta-coeff}.}
		\label{tab:triangulation-calibration}
		\begin{tabular}{ll}
			\hline 
			Beam offset & $s_p\simeq\SI{5.6}{mm}$ \\
			Beam deflection & $\alpha_p\simeq \ang{1.6}$ \\
			Calibration coefficients & $\beta_0 = \SI{133.8 \pm 0.2}{pixel}$ \\
			& $\beta_1 = \SI{3.69\pm 0.04}{pixel\per mm}$ \\ 
			\hline 
		\end{tabular} 
\end{table}

\subsection{Deposition height calculation}
\begin{figure}[ht]
	\centering
	\includegraphics[width=0.95\linewidth]{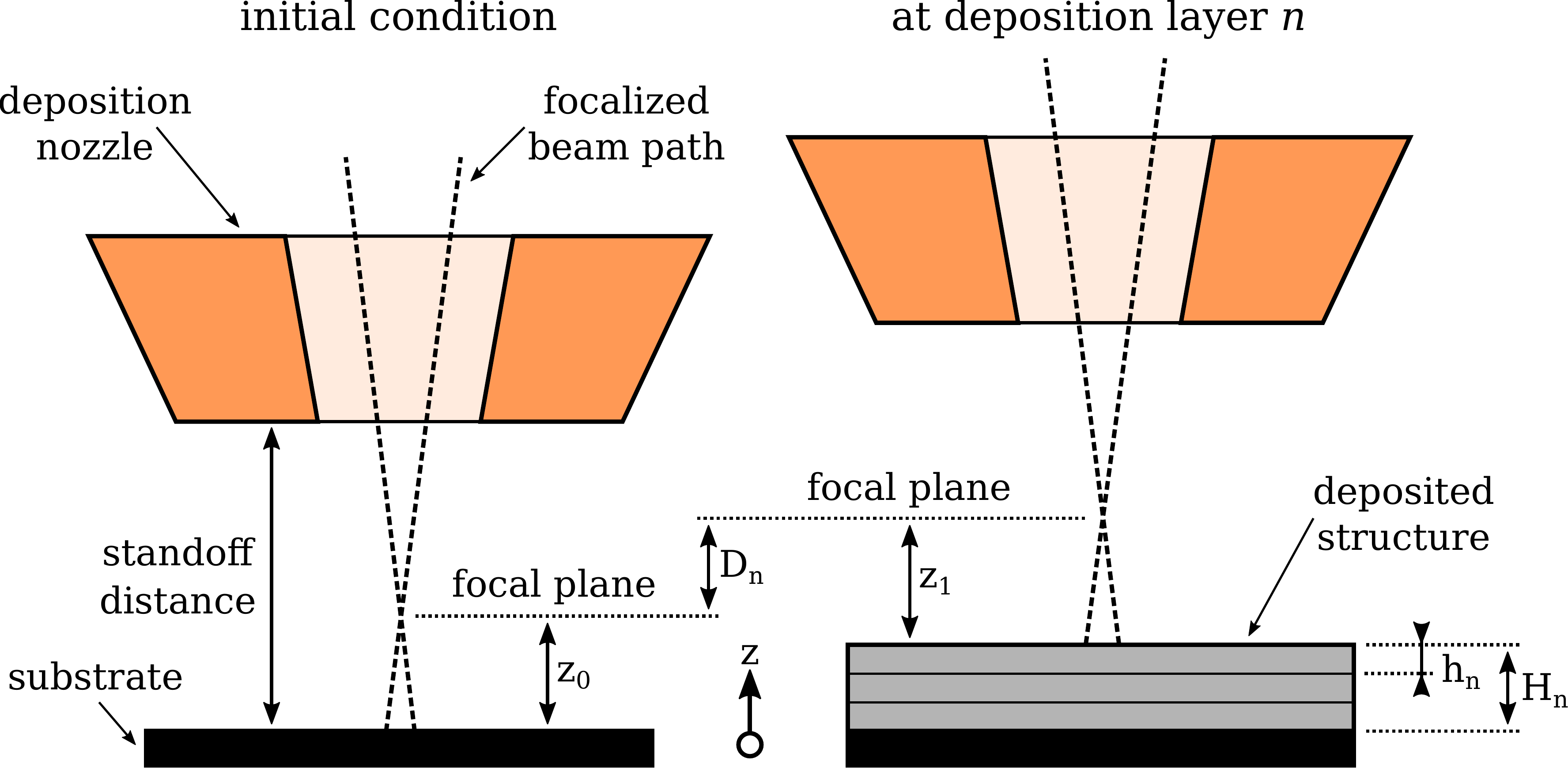}
	\caption{Sketch of the powder nozzle with the main dimensions related to the deposition process, both before (left) and after (right) the deposition of few layers and the consequent height increment of the deposition head.}
	\label{fig:configuration-ded}
\end{figure}

The deposition starts from the condition where the powder nozzle tip is positioned at the reference SOD from the substrate, as sketched in Fig.~\ref{fig:configuration-ded}. Once the process proceeds, the deposited structure grows layer by layer, and the vertical position of the deposition head is incremented accordingly. If $H(t)$ is the physical height of the deposited structure and $D(t)$ is the programmed robot height at the instant $t$, the height mismatch $D(t)-H(t)$ is equal to the relative height $\Delta z(t)$ measured with the procedure described previously, thus
\begin{equation}
\label{eq:height-mismatch-t}
D(t)-H(t) = z_1(t)-z_0  = \Delta z(t)\,,
\end{equation}
with
\begin{equation}
z_0 = z_1(t=0)\,.
\end{equation}

Considering a generic multi-layer structure, it is convenient to introduce a discrete notation for the geometrical variables related to the layer number $n$. Therefore, while depositing with a periodic track, $\Delta z_n$ can be defined as $\Delta z(t)$ at multiples of the track duration $T$:
\begin{equation}
\Delta z_n = \Delta z(nT) = z_1(nT)-z_1(0)\,.
\end{equation}
Analogously, in the assumption of a deterministic control of the robot coordinates, the total robot height $D_n$ is equal to the fixed height increment $d$ multiplied by the layer number $n$:
\begin{equation}
\label{eq:robot-height}
D_n = n d\,.
\end{equation}

The physical height $H_n$ of the deposited structure after $n$ layers is the sum of the layer thickness $h_j$, with $j$ ranging from $1$ to $n$, and it can be found from Eq.~\eqref{eq:height-mismatch-t} as
\begin{equation}
\label{eq:deposition-height}
H_n = \sum_{j=1}^{n} h_j = nd-\Delta z_n\,.
\end{equation}
The thickness of each deposited layer can be calculated as the difference of the deposition height between two consecutive layers:
\begin{equation}
\label{eq:calc-thickness}
h_{n} = H_{n}-H_{n-1} = d+ \Delta z_{n-1} - \Delta z_n\,.
\end{equation}
Consequently, the thickness mismatch, i.e. the error relative to the nominal height increment, is
\begin{equation}
h_{n} - d = \Delta z_{n-1} - \Delta z_n\,.
\end{equation}

\section{Height monitoring system applied to LMD}
\subsection{Cylinder deposition}
The device for the deposition height monitoring has been tested while building a single-wall hollow cylinder. The choice of such 3D geometry allowed to demonstrate the independence of the measurement working principle on the deposition direction. In fact, the transverse robot movement is, point by point, tangent to a circular path, hence all the directions are continuously probed.

The deposition has been performed using the process parameters reported in Table~\ref{tab:process-parameters}, which, from preliminary empirical tests, are known to produce good results in terms of quality and stability of the deposition. The process laser is operated at \SI{400}{W}, and its spot diameter on the substrate is set to \SI{1.4}{mm} by adjusting the position of the fiber collimation lens. The deposited material is AISI 316L stainless steel powder (\textsc{LPW}), having a grain size distribution between $\mathrm{D}_{10}=\SI{45}{\micro\meter}$ and $\mathrm{D}_{90}=\SI{90}{\micro\meter}$. The powder mass flow rate is set to \SI{9.2}{g \per \minute}. The substrate material is an AISI 304 plate, \SI{3}{mm} thick.

\begin{table}[ht]
	\centering
	\caption{Process parameters for the cylinder deposition.}
	\label{tab:process-parameters}
		\begin{tabular}{ll}
			\hline 
			Laser power & \SI{400}{W} \\ 
			Beam spot diameter & \SI{1.4}{mm} \\ 
			Deposited powder & AISI 316L \\ 
			Powder mass flow rate & \SI{9.2}{g \per \minute} \\ 
			Deposition velocity & \SI{20}{mm/s} \\
			Initial substrate SOD & \SI{12}{mm} \\ 
			Height increment & $d = \SI{0.2}{mm}$ \\
			Cylinder diameter & \SI{35}{mm} \\
			Layer number & $380$ \\
			\hline 
		\end{tabular} 
\end{table}

\begin{figure}[ht]
	\centering
	\includegraphics[width=0.55\linewidth]{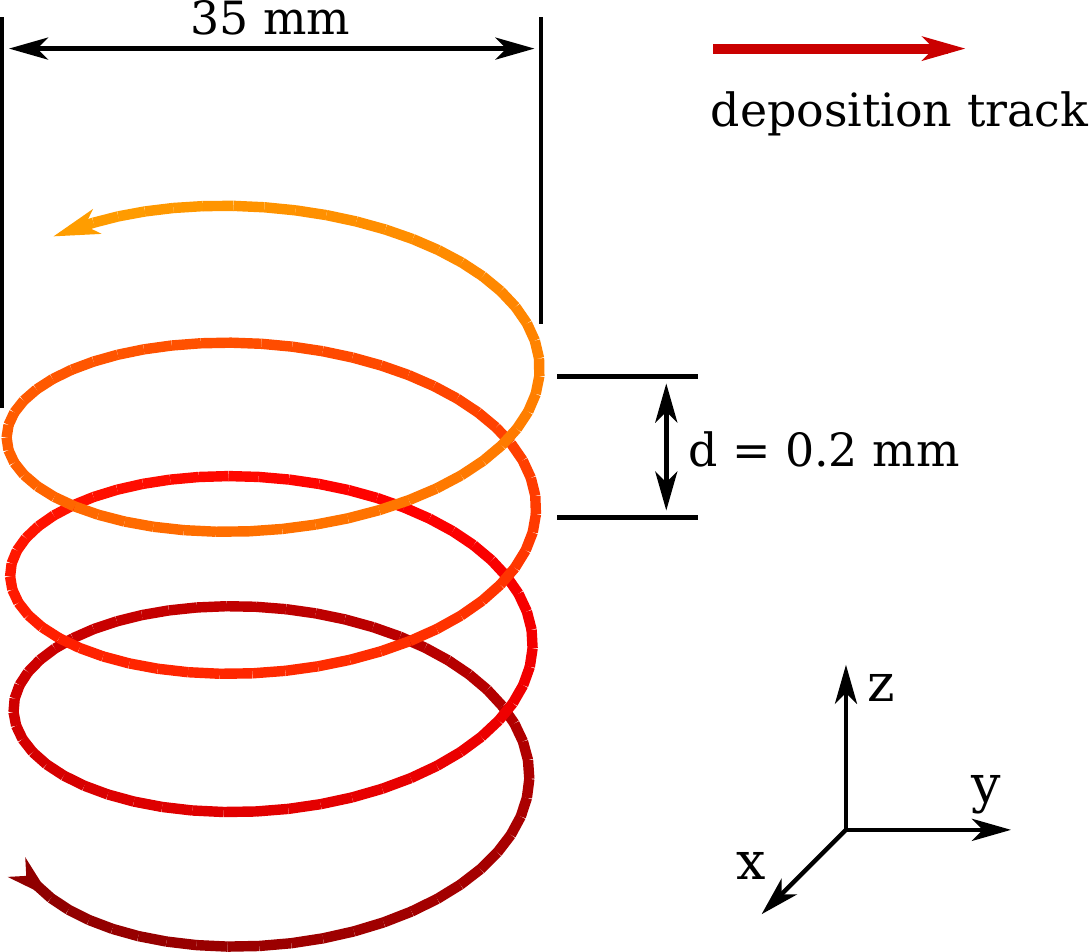}
	\caption{Scheme of the helicoidal robot path for the cylinder deposition (not in scale).}
	\label{fig:deposition-helix}
\end{figure}

The programmed robot path is a helix, as sketched in Fig.~\ref{fig:deposition-helix}. The multi-layer wall of the cylinder is built up by depositing the stainless steel powder on a \SI{35}{mm} diameter. The robot height is incremented continuously along the helix path, with a pinch height increment equal to $d = \SI{0.2}{mm}$ at each turn. The nominal linear deposition velocity is \SI{20}{mm/s}, for a total of about $380$ layers.

\subsection{Deposition height monitoring}
\begin{figure}[ht]
	\centering
	\includegraphics[width=0.95\linewidth]{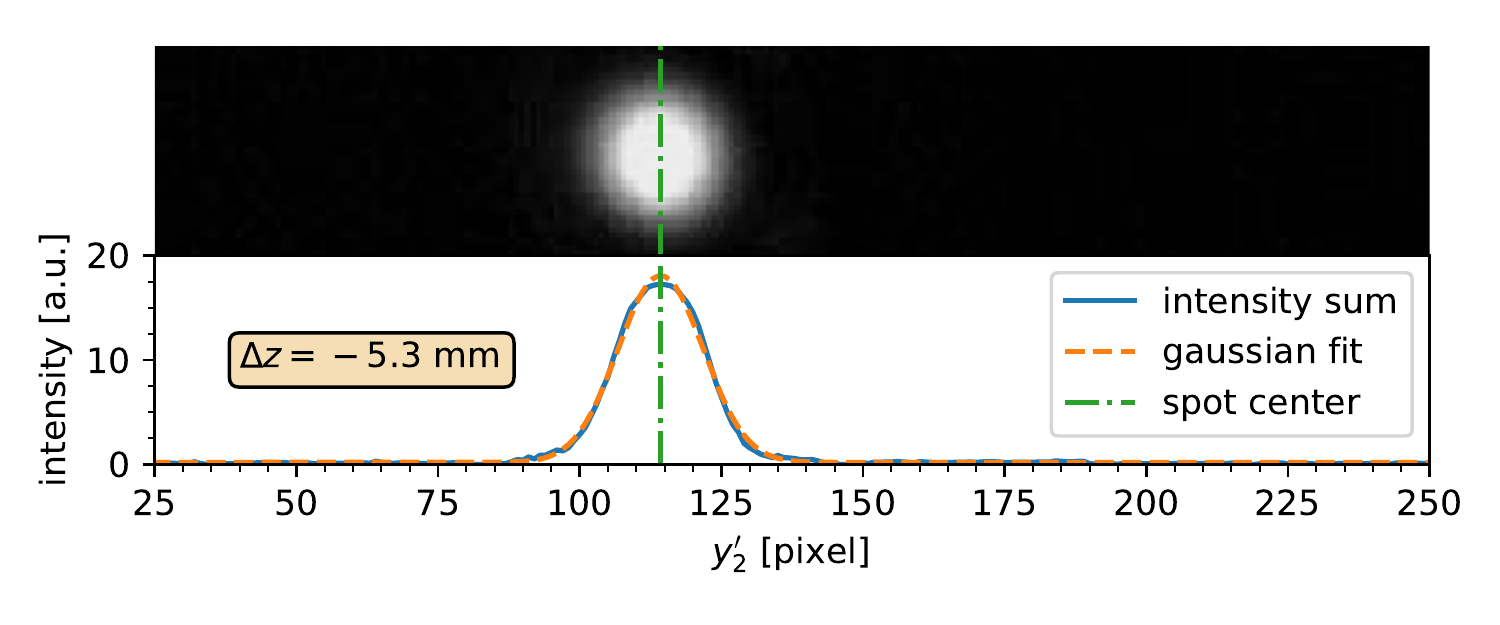}\\
	\includegraphics[width=0.95\linewidth]{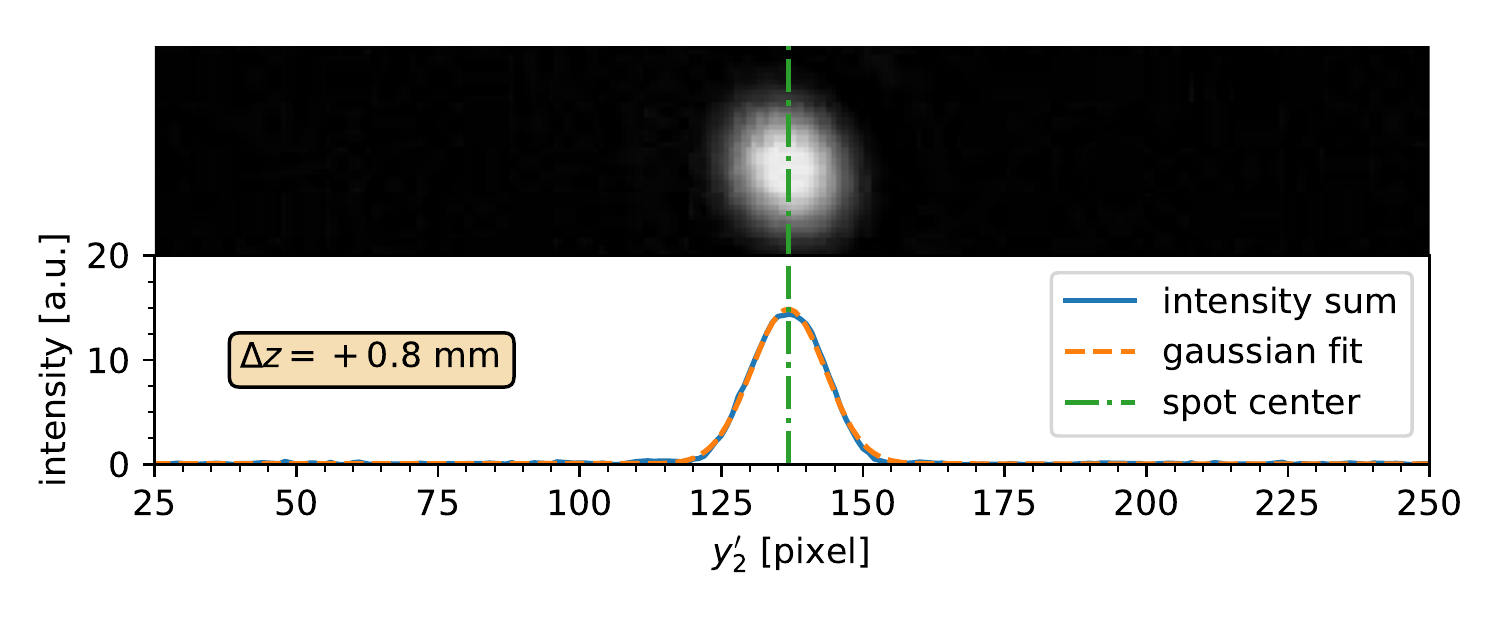}
	\caption{Two examples of probe spot images, acquired with the camera at different instants of the measurement data set (upper parts). The CCD column intensity is integrated and fitted to a Gaussian function to find the spot center and extract the height variation (lower parts).}
	\label{fig:frame-examples}
\end{figure}

Following Eq.~\eqref{eq:deposition-height}, the mismatch between the robot height $D$ and the measured height $H$ is equal to $\Delta z$, calculated using the inverse calibration expression of Eq.~\eqref{eq:delta-zeta-coeff}:
\begin{equation}
D(t)-H(t) = \Delta z(t) = \frac{y_2'(t)-\beta_0}{\beta_1}\,.
\end{equation}
Examples of frames acquired during the deposition are illustrated in Fig.~\ref{fig:frame-examples}, reporting the image of the probe spot on the deposition area and the respective height mismatch value, calculated from the Gaussian function fitting procedure on the image intensity. The raw signal $\Delta z(t)$ is smoothed with moving average over a period of $4$ frames, i.e., \SI{40}{ms}, in order to reduce the measurement shot noise. Eventual missing or invalid values are linearly interpolated.

\begin{figure}[ht]
	\centering
	\includegraphics[width=0.9\linewidth]{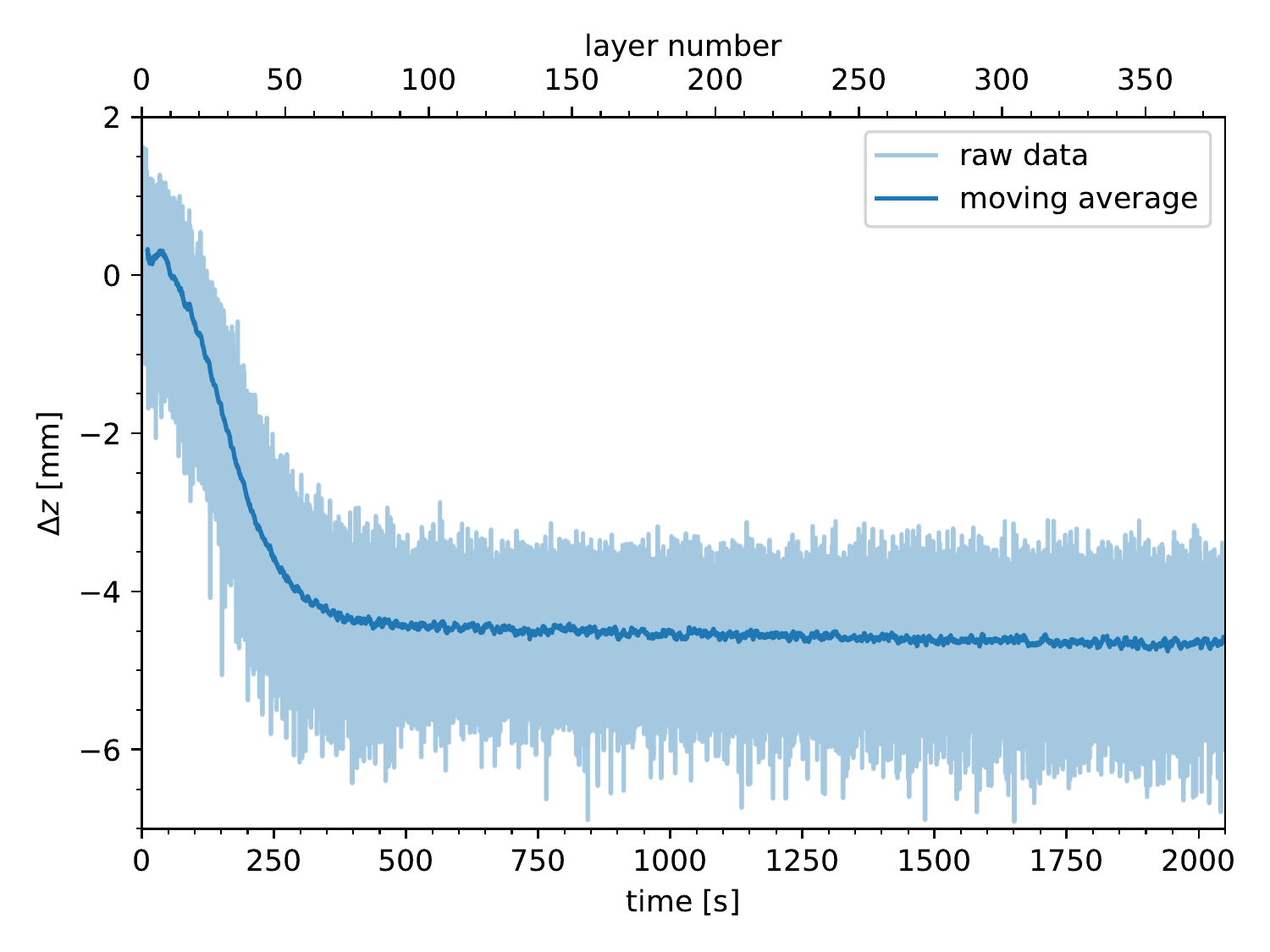}
	\caption{Mismatch between programmed robot height and measured deposition height, plotted as a function of deposition time. The corresponding layer number scale is also reported. Both the raw data and the moving average over a period of $1$ layer are reported.}
	\label{fig:standoff-time}
\end{figure}

In the case of accordance among deposition growth rate and robot path height increment, the mismatch $\Delta z(t)$ should remain equal to zero, meaning that the relative deposition height does not change from to its initial value. However, as it can be observed from Fig.~\ref{fig:standoff-time}, the behavior of the measured $\Delta z(t)$ shows an initial transient in the actual deposition height, which brings to a final mismatch of few millimeters. The negative sign of $\Delta z(t)$ means that the process grows faster than expected during the initial layers, with a reduction of the SOD and a final structure that is higher than expected.

\begin{figure}[ht]
	\centering
	\includegraphics[width=0.9\linewidth]{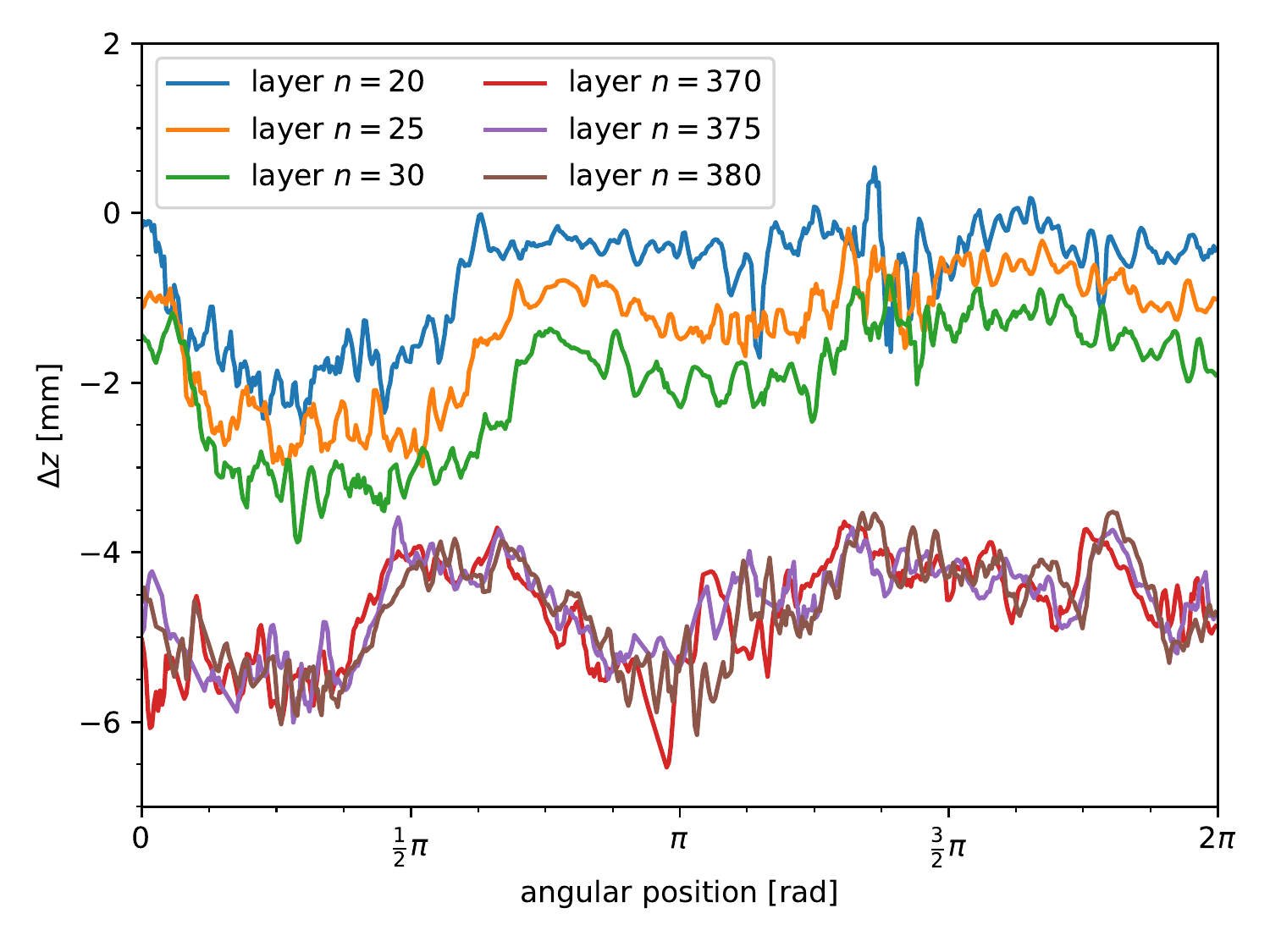}
	\caption{Height mismatch measured for few example layers, taken both in the initial transient and in the final stabilized intervals of the process growth. The layer profile is plotted as a function of the cylinder angular coordinate. The deposition defects introduce oscillations which propagates layer to layer.}
	\label{fig:standoff-zoom}
\end{figure}

The oscillations in the raw signal of Fig.~\ref{fig:standoff-time} around its average trend are given by several factors, such as the intrinsic measurement noise, the robot vibrations, and the presence of deposition defects. The signal for few deposition layer examples is reported in Fig.~\ref{fig:standoff-zoom}, translated as a function of the angular coordinate of the cylinder deposition track. As a first observation, an increasing offset in $\Delta z$ can be observed between the subsequent layers chosen in the initial transient interval of the deposition. Conversely, at the end of the deposition process the behavior of $\Delta z$ between subsequent layers stabilizes, with the deposition defects propagating layer to layer and introducing height ripples of the order of $1$--\SI{2}{mm}  within each helix turn.

\begin{figure}[hpt]
	\centering
	\includegraphics[width=0.85\linewidth]{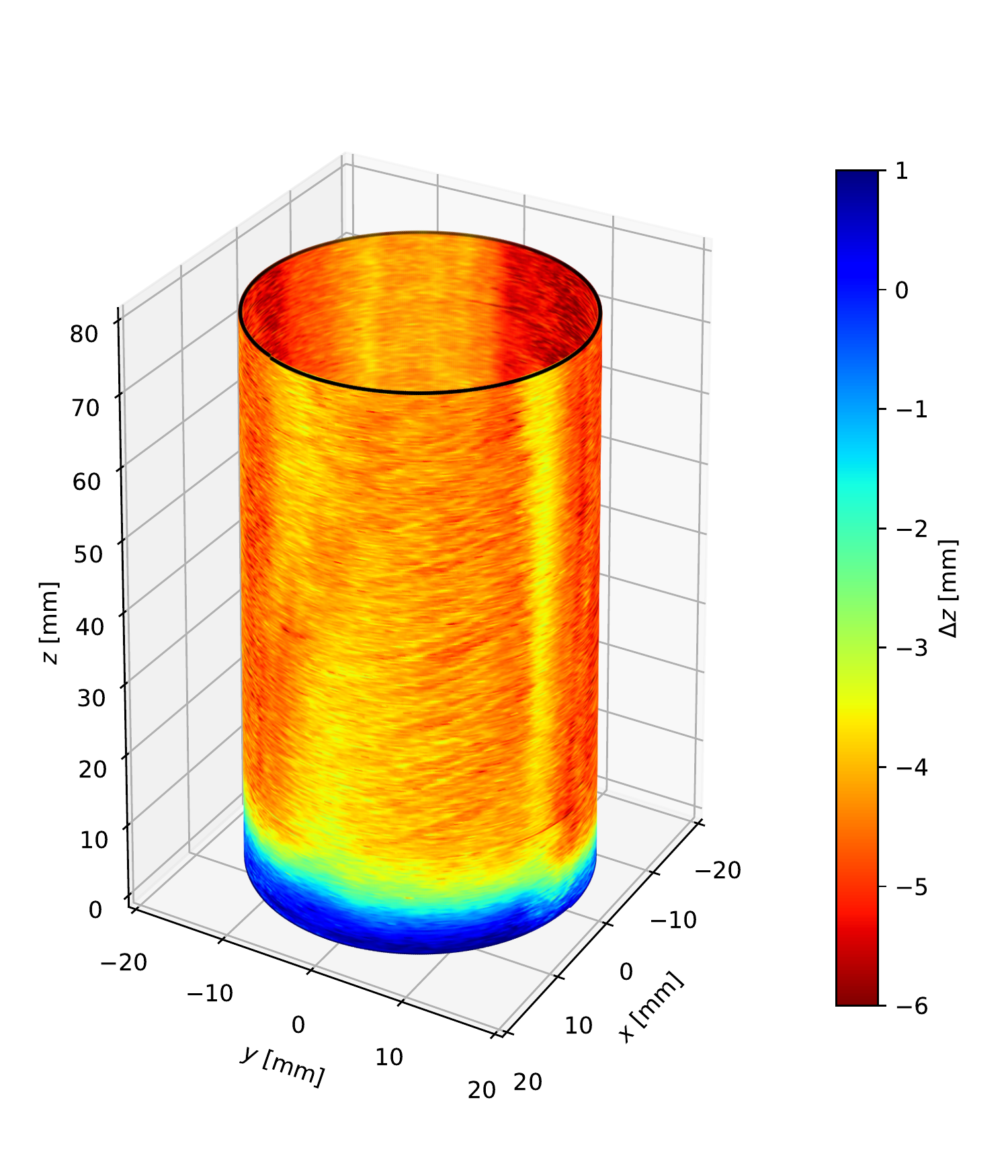}
	\caption{Mismatch between programmed robot height and measured cylinder height, in colormap representation and plotted as a function of robot coordinates.}
	\label{fig:cylinder3d-height-diff}
\end{figure}

\begin{figure}[ht]
	\centering
	\includegraphics[width=0.5\linewidth]{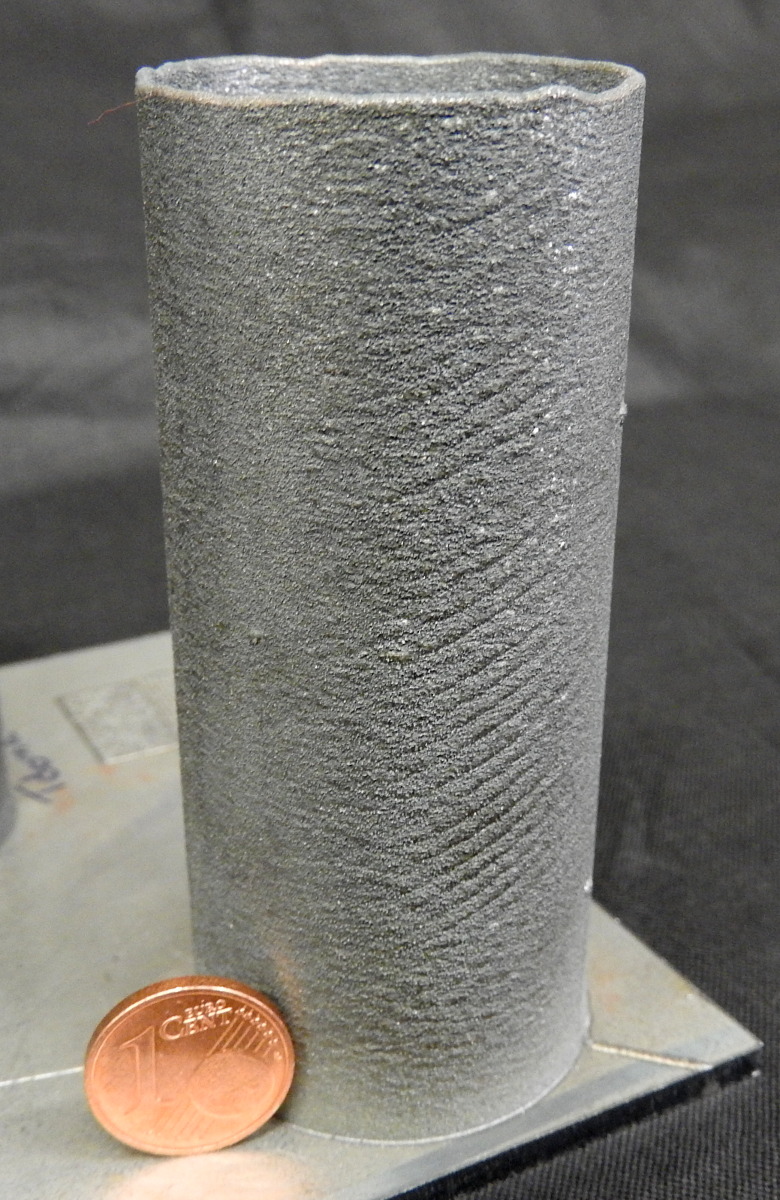}
	\caption{Photo of the deposited cylinder.}
	\label{fig:img-cylinder}
\end{figure}

In order to better visualize the height variations along the deposition track, the temporal sequence of the height measurement can be be translated into space coordinates knowing the camera acquisition frame rate and the deposition velocity. This allows to obtain a non-destructive inspection of the deposited layers along the vertical direction. Accordingly, the height mismatch $\Delta z$ reported in Fig.~\ref{fig:cylinder3d-height-diff} is plotted in false colors as a function of the coordinates referred to the helix center on the substrate and calculated from the programmed robot path. Stripe-like defects arising during the deposition process can be clearly recognized with this kind of 3D representation, finding visual correspondences in the actual deposited cylinder illustrated in Fig.~\ref{fig:img-cylinder}. Although the possibility of a dimensional comparison between the inline measurement and the final structure is limited by effects such as thermal expansion or presence of porosities, such qualitative considerations highlight that deposition defects and lateral roughness can be also correlated to local height variations.

\subsection{Self-regulation of the layer thickness}
\begin{figure}[ht]
	\centering
	\includegraphics[width=0.9\linewidth]{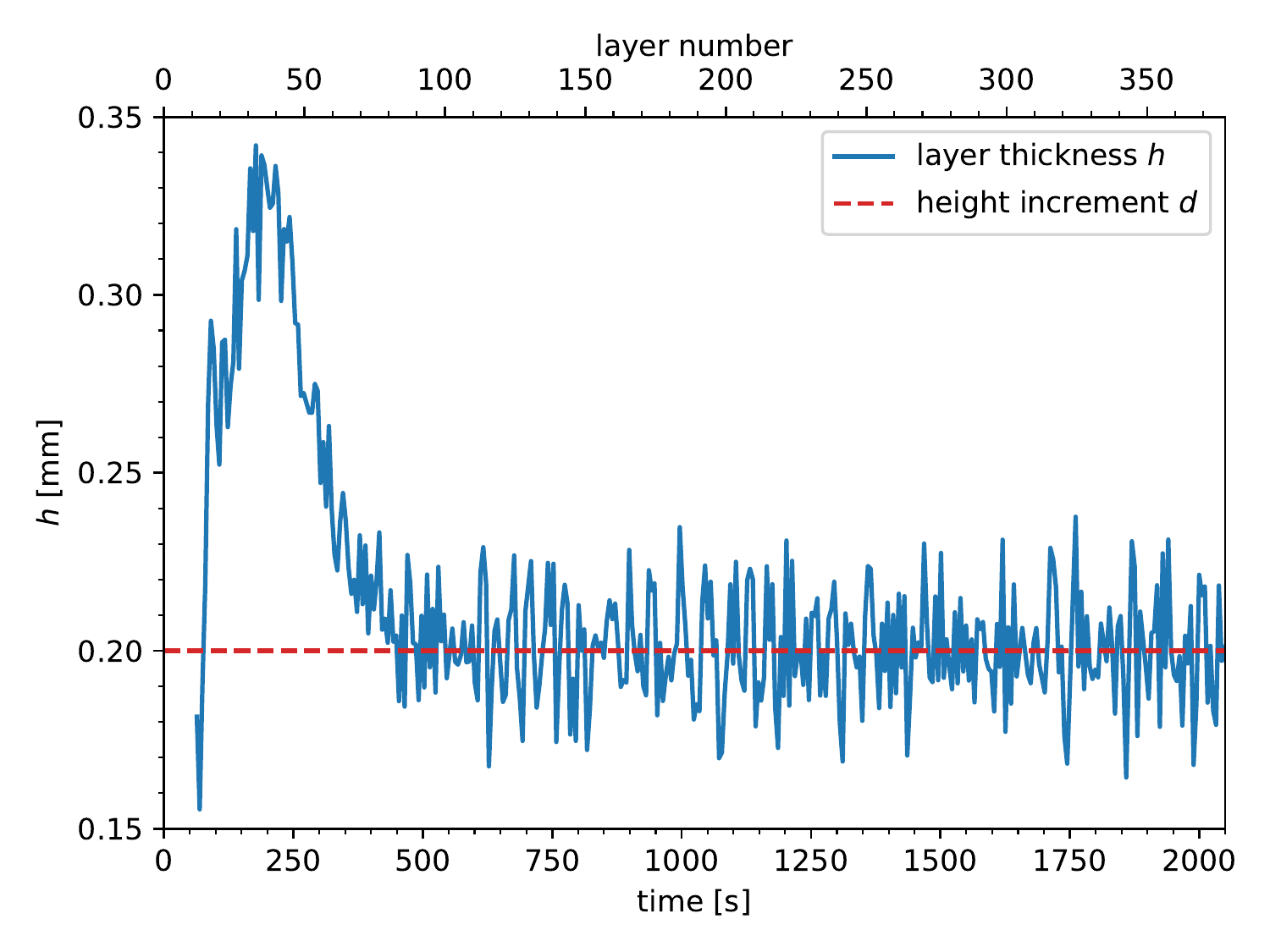}
	\caption{Layer thickness calculated from the height measurement, plotted as a function of deposition time and smoothed with a moving average over $3$ layer periods. The fixed robot height increment is reported for comparison.}
	\label{fig:tickness-time}
\end{figure}

\begin{figure}[hpt]
	\centering
	\includegraphics[width=0.85\linewidth]{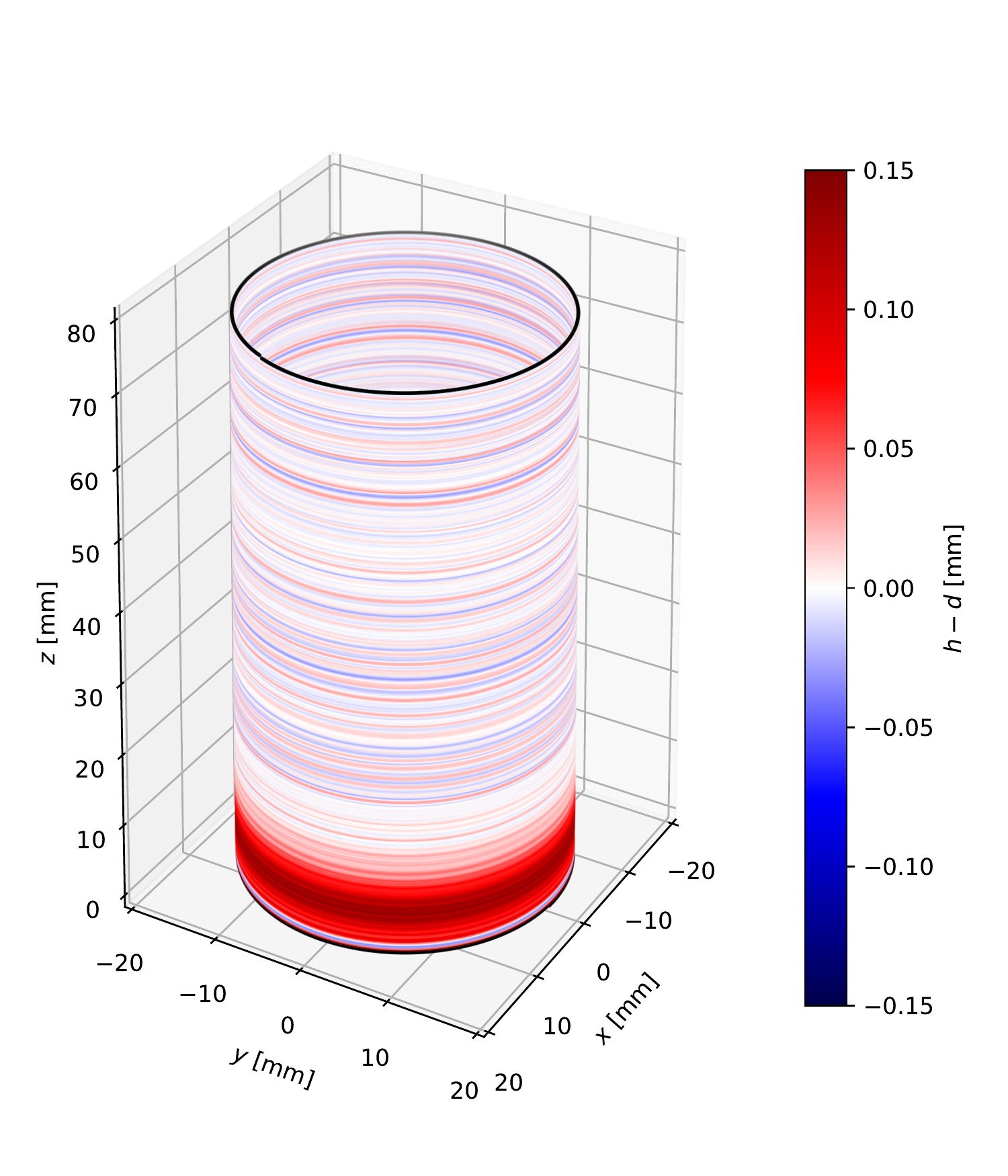}
	\caption{Mismatch between the measured layer thickness and the robot height increment, in colormap representation and plotted as a function of robot coordinates. The data are smoothed with a moving average over $3$ layer periods.}
	\label{fig:cylinder3d-layer-diff}
\end{figure}

The layer thickness measured during the deposition process is plotted in Fig.~\ref{fig:tickness-time}, calculated with Eq.~\eqref{eq:calc-thickness} from the height averaged over each deposition turn. As expected from the transient already observed while commenting the results of Fig.~\ref{fig:standoff-time}, a faster deposition growth during the initial interval reflects into thicker layers, whose thickness $h$ departs from the robot height increment $d=\SI{0.2}{mm}$ before converging to a stationary condition. At its maximum deviation, $h$ exceeds $d$ by the $75\%$. A spatial representation of the layer thickness mismatch $h-d$ is reported in Fig.~\ref{fig:cylinder3d-layer-diff}, plotted in false colors as a function of the deposition coordinates similarly to Fig.~\ref{fig:cylinder3d-height-diff}.

Although a deeper experimental study should be necessary for a precise interpretation, the evolution observed in the deposition growth can be explained qualitatively in terms of a self-regulating mechanism of the powder deposition process, similarly to what has been observed in other works \cite{zhu_influence_2011,zhu_influence_2012}. Two phenomena with contrast effects can be hypothesized. Firstly, it should be considered that size and temperature variations of the melt pool lead to changes in the powder catchment efficiency \cite{pinkerton_significance_2004}, hence in the thickness of the deposited layer. Specifically, an increase in the melt pool temperature is typically observed during the initial deposition layers before stabilizing \cite{bi_development_2007,manvatkar_estimation_2011}. This can be explained in terms of balance between the energy carried by the laser beam and the limited thermal conduction along the growing structure, with the substrate acting as heat sink. Such condition supports the melting and deposition of the powder particles on a wider and hotter melt pool, explaining the layer thickness growth observed at the beginning of the process.

The second phenomenon that must be considered happens if the structure grows faster than the robot height, hence the standoff distance between nozzle tip and the workpiece decreases, reflecting into a lower efficiency of the powder-laser interaction. In fact, if the SOD becomes too low, the deposition area goes out from the optimal working field of the powder cone, with the remelting process being predominant when the powder flow and the laser beam do not overlap efficiently on the substrate. With a reduced deposition efficiency the SOD increases again, hence tending back to a better deposition condition. Such cyclical phenomenon happens all over the process, with a compensation between the deposition efficiency and the layer thickness growth. The overall result leads to a self-regulating mechanism, with an asymptotic tendency of the growth rate to an equilibrium condition, corresponding to the minimum SOD permitted by the powder nozzle configuration.

The study of such kinds of mechanisms may allow for a rigorous identification of the optimal process conditions, such as the initial standoff and the programmed height increment, which can be adjusted to minimize the initial growth transient and to accelerate the convergence to the self-regulating regime in a passive way \cite{zhu_influence_2012}. Another approach may consider the integration of the height measurement with control systems for a dynamical adaption of the deposition parameters to the process evolution. This might be achieved by actively modulating the laser power for a fast feedback response \cite{song_control_2012,hofman_camera_2012}, the powder flow rate to compensate the variable catchment efficiency \cite{tang_layer--layer_2011}, or the deposition track coordinates and velocity to correct any dimensional mismatch \cite{fathi_clad_2007,sammons_repetitive_2018}.

\section{Conclusions}
The current work presented the design and implementation of a system for monitoring the deposition height on a LMD setup. The coaxial configuration of the triangulation measurement allows for flexibility in terms of deposition strategy, overcoming some of the off-axis method limits, such as blindness along specific directions. A measurement laser beam probes the melt pool height, sharing the same optical path of the process laser beam, without the need of significant modifications to the existing setup. The device has been demonstrated for monitoring the deposition of stainless steel powder for building a multi-layer cylinder, showing its robustness even in the presence of the process emission and of the metallic powder flow. Its operation can be generalized to more complex 3D geometries and to different materials, since the height measurement is direct instead of being linked to process models.

The actual thickness of the deposited layers has been calculated from the triangulation measurements. Although the system sensitivity might be improved with a higher imaging resolution, this was enough to highlight the emergence of a self-regulating mechanism in the deposition growth, which, after an initial transient, converges to the programmed height increment. Understanding such mechanism is important, since layer thickness variations can lead to shape irregularities and dimensional mismatches. The translation of the measurement into space coordinates allowed to obtain a 3D spatial map of the deposited structure, showing the formation of local defects which propagate along the growth direction.

A precise synchronization of the measurement with the robot movements would allow to use the system for monitoring in real-time the deposition growth, without the need of post-process analysis. Such possibility may represent a simple, low-cost, and non-intrusive solution for controlling the process on generic coaxial LMD setups, without the need of more complex and expensive devices. The measurement error signal might be employed in feedback control systems, allowing to adapt the process parameters to the actual deposition growth and to correct the emergence of defects, hence improving quality and regularity of the deposited structures.

\section*{Acknowledgments}
The authors acknowledge the technical support from BLM Group. Guendalina Catalano is acknowledged for her support during the experiments. This work was supported by European Union, Repubblica Italiana, Regione Lombardia and FESR for the project MADE4LO under the call ``POR FESR 2014-2020 ASSE I -- AZIONE I.1.B.1.3''.
	
%\section*{References}
\bibliography{bibliography}

\begin{thebibliography}{10}
\expandafter\ifx\csname url\endcsname\relax
  \def\url#1{\texttt{#1}}\fi
\expandafter\ifx\csname urlprefix\endcsname\relax\def\urlprefix{URL }\fi
\expandafter\ifx\csname href\endcsname\relax
  \def\href#1#2{#2} \def\path#1{#1}\fi

\bibitem{frazier_metal_2014}
W.~E. Frazier, Metal {{Additive Manufacturing}}: {{A Review}}, Journal of
  Materials Engineering and Performance 23~(6) (2014) 1917--1928.
\newblock \href {https://doi.org/10.1007/S11665-014-0958-Z}
  {\path{doi:10.1007/S11665-014-0958-Z}}.

\bibitem{mazumder_closed_2000}
J.~Mazumder, D.~Dutta, N.~Kikuchi, A.~Ghosh, Closed loop direct metal
  deposition: Art to part, Optics and Lasers in Engineering 34~(4) (2000)
  397--414.
\newblock \href {https://doi.org/10.1016/S0143-8166(00)00072-5}
  {\path{doi:10.1016/S0143-8166(00)00072-5}}.

\bibitem{purtonen_monitoring_2014}
T.~Purtonen, A.~Kalliosaari, A.~Salminen, Monitoring and {{Adaptive Control}}
  of {{Laser Processes}}, Physics Procedia 56 (2014) 1218--1231.
\newblock \href {https://doi.org/10.1016/J.PHPRO.2014.08.038}
  {\path{doi:10.1016/J.PHPRO.2014.08.038}}.

\bibitem{shamsaei_overview_2015}
N.~Shamsaei, A.~Yadollahi, L.~Bian, S.~M. Thompson, An overview of {{Direct
  Laser Deposition}} for additive manufacturing; {{Part II}}: {{Mechanical}}
  behavior, process parameter optimization and control, Additive Manufacturing
  8 (2015) 12--35.
\newblock \href {https://doi.org/10.1016/J.ADDMA.2015.07.002}
  {\path{doi:10.1016/J.ADDMA.2015.07.002}}.

\bibitem{sammons_repetitive_2018}
P.~M. Sammons, M.~L. Gegel, D.~A. Bristow, R.~G. Landers, Repetitive {{Process
  Control}} of {{Additive Manufacturing With Application}} to {{Laser Metal
  Deposition}}, IEEE Transactions on Control Systems Technology (2018)
  1--10\href {https://doi.org/10.1109/TCST.2017.2781653}
  {\path{doi:10.1109/TCST.2017.2781653}}.

\bibitem{meriaudeau_control_1996}
F.~Meriaudeau, F.~Truchetet, Control and optimization of the laser cladding
  process using matrix cameras and image processing, Journal of Laser
  Applications 8~(6) (1996) 317--324.
\newblock \href {https://doi.org/10.2351/1.4745438}
  {\path{doi:10.2351/1.4745438}}.

\bibitem{hu_sensing_2003}
D.~Hu, R.~Kovacevic, Sensing, modeling and control for laser-based additive
  manufacturing, International Journal of Machine Tools and Manufacture 43~(1)
  (2003) 51--60.
\newblock \href {https://doi.org/10.1016/S0890-6955(02)00163-3}
  {\path{doi:10.1016/S0890-6955(02)00163-3}}.

\bibitem{hofman_camera_2012}
J.~T. Hofman, B.~Pathiraj, J.~{van Dijk}, D.~F. {de Lange}, J.~Meijer, A camera
  based feedback control strategy for the laser cladding process, Journal of
  Materials Processing Technology 212~(11) (2012) 2455--2462.
\newblock \href {https://doi.org/10.1016/J.JMATPROTEC.2012.06.027}
  {\path{doi:10.1016/J.JMATPROTEC.2012.06.027}}.

\bibitem{tapia_review_2014}
G.~Tapia, A.~Elwany, A {{Review}} on {{Process Monitoring}} and {{Control}} in
  {{Metal}}-{{Based Additive Manufacturing}}, Journal of Manufacturing Science
  and Engineering 136~(6) (2014) 060801--060801--10.
\newblock \href {https://doi.org/10.1115/1.4028540}
  {\path{doi:10.1115/1.4028540}}.

\bibitem{everton_review_2016}
S.~K. Everton, M.~Hirsch, P.~Stravroulakis, R.~K. Leach, A.~T. Clare, Review of
  in-situ process monitoring and in-situ metrology for metal additive
  manufacturing, Materials \& Design 95 (2016) 431--445.
\newblock \href {https://doi.org/10.1016/J.MATDES.2016.01.099}
  {\path{doi:10.1016/J.MATDES.2016.01.099}}.

\bibitem{kim_review_2018}
H.~Kim, Y.~Lin, T.-L.~B. Tseng, A review on quality control in additive
  manufacturing, Rapid Prototyping Journal 24~(3) (2018) 645--669.
\newblock \href {https://doi.org/10.1108/RPJ-03-2017-0048}
  {\path{doi:10.1108/RPJ-03-2017-0048}}.

\bibitem{peyre_analytical_2008}
P.~Peyre, P.~Aubry, R.~Fabbro, R.~Neveu, A.~Longuet, Analytical and numerical
  modelling of the direct metal deposition laser process, Journal of Physics D:
  Applied Physics 41~(2) (2008) 025403.
\newblock \href {https://doi.org/10.1088/0022-3727/41/2/025403}
  {\path{doi:10.1088/0022-3727/41/2/025403}}.

\bibitem{kovalev_theoretical_2011}
O.~B. Kovalev, A.~V. Zaitsev, D.~Novichenko, I.~Smurov, Theoretical and
  {{Experimental Investigation}} of {{Gas Flows}}, {{Powder Transport}} and
  {{Heating}} in {{Coaxial Laser Direct Metal Deposition}} ({{DMD}})
  {{Process}}, Journal of Thermal Spray Technology 20~(3) (2011) 465--478.
\newblock \href {https://doi.org/10.1007/S11666-010-9539-3}
  {\path{doi:10.1007/S11666-010-9539-3}}.

\bibitem{pinkerton_significance_2004}
A.~J. Pinkerton, L.~Li, The significance of deposition point standoff
  variations in multiple-layer coaxial laser cladding (coaxial cladding
  standoff effects), International Journal of Machine Tools and Manufacture
  44~(6) (2004) 573--584.
\newblock \href {https://doi.org/10.1016/J.IJMACHTOOLS.2004.01.001}
  {\path{doi:10.1016/J.IJMACHTOOLS.2004.01.001}}.

\bibitem{zhu_influence_2012}
G.~Zhu, D.~Li, A.~Zhang, G.~Pi, Y.~Tang, The influence of laser and powder
  defocusing characteristics on the surface quality in laser direct metal
  deposition, Optics \& Laser Technology 44~(2) (2012) 349--356.
\newblock \href {https://doi.org/10.1016/J.OPTLASTEC.2011.07.013}
  {\path{doi:10.1016/J.OPTLASTEC.2011.07.013}}.

\bibitem{labudovic_three_2003}
M.~Labudovic, D.~Hu, R.~Kovacevic, A three dimensional model for direct laser
  metal powder deposition and rapid prototyping, Journal of Materials Science
  38~(1) (2003) 35--49.
\newblock \href {https://doi.org/10.1023/A:1021153513925}
  {\path{doi:10.1023/A:1021153513925}}.

\bibitem{amine_investigation_2014}
T.~Amine, J.~W. Newkirk, F.~Liou, Investigation of effect of process parameters
  on multilayer builds by direct metal deposition, Applied Thermal Engineering
  73~(1) (2014) 500--511.
\newblock \href {https://doi.org/10.1016/J.APPLTHERMALENG.2014.08.005}
  {\path{doi:10.1016/J.APPLTHERMALENG.2014.08.005}}.

\bibitem{manvatkar_spatial_2015}
V.~Manvatkar, A.~De, T.~DebRoy, Spatial variation of melt pool geometry, peak
  temperature and solidification parameters during laser assisted additive
  manufacturing process, Materials Science and Technology 31~(8) (2015)
  924--930.
\newblock \href {https://doi.org/10.1179/1743284714Y.0000000701}
  {\path{doi:10.1179/1743284714Y.0000000701}}.

\bibitem{fathi_clad_2007}
A.~Fathi, A.~Khajepour, E.~Toyserkani, M.~Durali, Clad height control in laser
  solid freeform fabrication using a feedforward {{PID}} controller, The
  International Journal of Advanced Manufacturing Technology 35~(3-4) (2007)
  280--292.
\newblock \href {https://doi.org/10.1007/S00170-006-0721-1}
  {\path{doi:10.1007/S00170-006-0721-1}}.

\bibitem{bi_development_2007}
G.~Bi, B.~Sch\"urmann, A.~Gasser, K.~Wissenbach, R.~Poprawe, Development and
  qualification of a novel laser-cladding head with integrated sensors,
  International Journal of Machine Tools and Manufacture 47~(3) (2007)
  555--561.
\newblock \href {https://doi.org/10.1016/J.IJMACHTOOLS.2006.05.010}
  {\path{doi:10.1016/J.IJMACHTOOLS.2006.05.010}}.

\bibitem{hua_feedback_2005}
Y.~Hua, J.~Choi, Feedback control effects on dimensions and defects of {{H13}}
  tool steel by direct metal deposition process, Journal of Laser Applications
  17~(2) (2005) 118--126.
\newblock \href {https://doi.org/10.2351/1.1848530}
  {\path{doi:10.2351/1.1848530}}.

\bibitem{asselin_development_2005}
M.~Asselin, E.~Toyserkani, M.~{Iravani-Tabrizipour}, A.~Khajepour, Development
  of trinocular {{CCD}}-based optical detector for real-time monitoring of
  laser cladding, in: {{IEEE International Conference Mechatronics}} and
  {{Automation}}, 2005, Vol.~3, 2005, pp. 1190--1196 Vol. 3.
\newblock \href {https://doi.org/10.1109/ICMA.2005.1626722}
  {\path{doi:10.1109/ICMA.2005.1626722}}.

\bibitem{iravani-tabrizipour_image-based_2007}
M.~{Iravani-Tabrizipour}, E.~Toyserkani, An image-based feature tracking
  algorithm for real-time measurement of clad height, Machine Vision and
  Applications 18~(6) (2007) 343--354.
\newblock \href {https://doi.org/10.1007/S00138-006-0066-7}
  {\path{doi:10.1007/S00138-006-0066-7}}.

\bibitem{song_control_2012}
L.~Song, V.~{Bagavath-Singh}, B.~Dutta, J.~Mazumder, Control of melt pool
  temperature and deposition height during direct metal deposition process, The
  International Journal of Advanced Manufacturing Technology 58~(1-4) (2012)
  247--256.
\newblock \href {https://doi.org/10.1007/S00170-011-3395-2}
  {\path{doi:10.1007/S00170-011-3395-2}}.

\bibitem{biegler_-situ_2018}
M.~Biegler, B.~Graf, M.~Rethmeier, In-situ distortions in {{LMD}} additive
  manufacturing walls can be measured with digital image correlation and
  predicted using numerical simulations, Additive Manufacturing 20 (2018)
  101--110.
\newblock \href {https://doi.org/10.1016/J.ADDMA.2017.12.007}
  {\path{doi:10.1016/J.ADDMA.2017.12.007}}.

\bibitem{heralic_increased_2010}
A.~Herali\'c, A.-K. Christiansson, M.~Ottosson, B.~Lennartson, Increased
  stability in laser metal wire deposition through feedback from optical
  measurements, Optics and Lasers in Engineering 48~(4) (2010) 478--485.
\newblock \href {https://doi.org/10.1016/J.OPTLASENG.2009.08.012}
  {\path{doi:10.1016/J.OPTLASENG.2009.08.012}}.

\bibitem{tang_layer--layer_2011}
L.~Tang, R.~G. Landers, Layer-to-{{Layer Height Control}} for {{Laser Metal
  Deposition Process}}, Journal of Manufacturing Science and Engineering
  133~(2) (2011) 021009.
\newblock \href {https://doi.org/10.1115/1.4003691}
  {\path{doi:10.1115/1.4003691}}.

\bibitem{denlinger_effect_2015}
E.~R. Denlinger, J.~C. Heigel, P.~Michaleris, T.~A. Palmer, Effect of
  inter-layer dwell time on distortion and residual stress in additive
  manufacturing of titanium and nickel alloys, Journal of Materials Processing
  Technology 215 (2015) 123--131.
\newblock \href {https://doi.org/10.1016/J.JMATPROTEC.2014.07.030}
  {\path{doi:10.1016/J.JMATPROTEC.2014.07.030}}.

\bibitem{heigel_situ_2015}
J.~C. Heigel, P.~Michaleris, T.~A. Palmer, In situ monitoring and
  characterization of distortion during laser cladding of
  {{Inconel}}\textregistered{} 625, Journal of Materials Processing Technology
  220 (2015) 135--145.
\newblock \href {https://doi.org/10.1016/J.JMATPROTEC.2014.12.029}
  {\path{doi:10.1016/J.JMATPROTEC.2014.12.029}}.

\bibitem{segerstark_investigation_2017}
A.~Segerstark, J.~Andersson, L.-E. Svensson, Investigation of laser metal
  deposited {{Alloy}} 718 onto an {{EN}} 1.4401 stainless steel substrate,
  Optics \& Laser Technology 97 (2017) 144--153.
\newblock \href {https://doi.org/10.1016/J.OPTLASTEC.2017.05.038}
  {\path{doi:10.1016/J.OPTLASTEC.2017.05.038}}.

\bibitem{heralic_height_2012}
A.~Herali\'c, A.-K. Christiansson, B.~Lennartson, Height control of laser
  metal-wire deposition based on iterative learning control and {{3D}}
  scanning, Optics and Lasers in Engineering 50~(9) (2012) 1230--1241.
\newblock \href {https://doi.org/10.1016/J.OPTLASENG.2012.03.016}
  {\path{doi:10.1016/J.OPTLASENG.2012.03.016}}.

\bibitem{donadello_coaxial_2018}
S.~Donadello, M.~Motta, A.~G. Demir, B.~Previtali, Coaxial laser triangulation
  for height monitoring in laser metal deposition, in: Procedia {{CIRP}},
  Vol.~74, 2018, pp. 144--148.
\newblock \href {https://doi.org/10.1016/j.procir.2018.08.066}
  {\path{doi:10.1016/j.procir.2018.08.066}}.

\bibitem{zhu_influence_2011}
G.~Zhu, D.~Li, A.~Zhang, G.~Pi, Y.~Tang, The influence of standoff variations
  on the forming accuracy in laser direct metal deposition, Rapid Prototyping
  Journal 17~(2) (2011) 98--106.
\newblock \href {https://doi.org/10.1108/13552541111113844}
  {\path{doi:10.1108/13552541111113844}}.

\bibitem{manvatkar_estimation_2011}
V.~D. Manvatkar, A.~A. Gokhale, G.~J. Reddy, A.~Venkataramana, A.~De,
  Estimation of {{Melt Pool Dimensions}}, {{Thermal Cycle}}, and {{Hardness
  Distribution}} in the {{Laser}}-{{Engineered Net Shaping Process}} of
  {{Austenitic Stainless Steel}}, Metallurgical and Materials Transactions A
  42~(13) (2011) 4080--4087.
\newblock \href {https://doi.org/10.1007/S11661-011-0787-8}
  {\path{doi:10.1007/S11661-011-0787-8}}.

\end{thebibliography}
\end{document}